\documentclass[aps,prd,twocolumn,superscriptaddress,nofootinbib,floatfix]{revtex4-2}

\IfFileExists{newtxtext.sty}{
  \usepackage[T1]{fontenc}
  \usepackage{newtxtext}
  \usepackage{newtxmath}
}{
  \usepackage{mathptmx}
}

\usepackage{amsmath,amssymb,bm}
\usepackage{graphicx}
\usepackage{booktabs}
\PassOptionsToPackage{hyphens}{url}
\usepackage[colorlinks=true,linkcolor=blue,citecolor=magenta,urlcolor=blue]{hyperref}
\AtBeginDocument{\let\hbar\hslash}

\def\doi#1{\href{https://doi.org/#1}{\color{blue}#1}}
\usepackage{tikz}
\usetikzlibrary{shapes.geometric,shapes.symbols}
\DeclareRobustCommand{\orcidicon}{
  \begin{tikzpicture}
    \definecolor{orcidgreen}{HTML}{166B3A}
    \draw[orcidgreen, fill=orcidgreen] (0,0)
    circle [radius=0.16]
    node[white] {{\fontfamily{qag}\selectfont \tiny ID}};
  \end{tikzpicture}
  \hspace{-2mm}
}

\foreach \x in {A, ..., Z}{
  \expandafter\xdef\csname orcid\x\endcsname{\noexpand\href{https://orcid.org/\csname orcidauthor\x\endcsname}{\noexpand\orcidicon}}
}

\newcommand{\NS}{{\rm NS}}

\newcommand{\GB}{Gauss--Bonnet}
\newcommand{\GW}{Gibbons--Werner}
\newcommand{\PPN}{PPN}

\begin{document}

\title{Screened Scalar Hair and the Weak-Lensing Separation of\\ Black Holes from Neutron Stars in Quadratic $f(R)$ Gravity}

\author{Yashmitha Kumaran\,\orcidB{}}
\affiliation{Instituto de Astronom\'ia, Universidad Nacional Aut\'onoma de M\'exico, \\ AP 70-264, Ciudad de M\'exico 04510, M\'exico.}

\author{Il\'idio Lopes\,\orcidA{}}
\affiliation{Centro de Astrof\'isica e Gravita\c{c}\~ao--CENTRA, Departamento de F\'isica, \\ Instituto Superior T\'ecnico--IST, Universidade de Lisboa--UL, \\ Av.~Rovisco Pais~1, 1049-001 Lisboa, Portugal}

\begin{abstract}
Quadratic $f(R)$ gravity carries a massive scalar degree of freedom, the scalaron, whose finite range $\lambda$ screens its influence on the geometry outside a compact object. We show that this screening severs the exterior of a black hole from that of a neutron star of the same mass. The correction to the Schwarzschild metric is Yukawa-suppressed, falling as $e^{-r/\lambda}$ instead of polynomially in the coupling, and is therefore invisible to any expansion in powers of that coupling; the exterior is moreover intrinsically isotropic, so that inverting the temporal potential alone, as one does for single-potential solutions, fails to solve the field equations. Applying the \GW\ \GB\ theorem to this geometry, we find the leading deflection angle to be exactly the general-relativistic $4GM/b$, with $b$ the impact parameter, the scalaron contributions cancelling identically in the combination that bends light. The first correction carries the unfamiliar signature $\lambda^{-1/2}b^{-3/2}e^{-b/\lambda}$, screened beyond the scalaron range and confirmed against direct quadrature to around $20\%$ at $b=4\lambda$, improving to $7\%$ at $b=12\lambda$. What survives is not a difference of degree but of kind. The ingredients that deliver it --- namely, non-analyticity in the coupling, the intrinsically isotropic gauge, the closed-form screened correction and the pressure-weighted scalar charge --- are obtained within a single, self-consistent derivation here. A static black hole carries no scalar hair, and lenses precisely as general relativity requires; a neutron star acquires a scalar charge weighted by the pressure supporting it against collapse, and does not. Weak lensing therefore closes as a discriminant of the theory, whilst the horizon, as against a material surface, remains one in principle. The observational advantage lies not in bending angles at large impact parameter but in the strong-field imaging of the photon sphere, and in the stellar interior, where the scalar charge is fixed by the equation of state.
\end{abstract}

\maketitle

\section{Introduction}
\label{sec:intro}

General relativity has withstood every weak-field test to which it has been subjected, and yet the strong-field regime near compact objects remains relatively unexplored. Higher-derivative extensions of the Einstein--Hilbert action arise naturally in low-energy effective field theories of quantum gravity and in inflationary cosmology, and amongst them quadratic $f(R)=R+\alpha R^2$ gravity, the Starobinsky model, occupies a distinguished position: it is the simplest curvature-squared extension, it is ghost-free, and it is already tightly constrained in cosmology and in the Solar System~\citep{Starobinsky1980,Sotiriou2010,DeFelice2010, NojiriOdintsov2011}. The same action has also been applied to compact stars directly, where its scalar degree of freedom leaves an imprint on stellar structure distinct from the exterior signatures considered here~\citep{Panotopoulos2018a,Panotopoulos2018b}.

A recurring difficulty in testing modified gravity with compact objects is parameter degeneracy: a coupling such as $\alpha$ can imprint itself on photon trajectories in ways that resemble the effect of an unknown equation of state, so that lensing or shadow observables alone rarely fix both simultaneously~\citep{Biswas2024}. It is natural to ask whether the nature of the object supplying the boundary, an event horizon as against a material stellar surface, breaks this degeneracy: a black hole and a neutron star of identical mass generate, in general relativity, identical exterior spacetimes down to the stellar surface, and identical light bending outside it. We ask whether the same holds in quadratic $f(R)$ gravity. Framed in terms of the scalar-tensor picture introduced below, the two cases are distinguished by whether the scalaron is sourced at all: a black hole exterior is pure vacuum, with $T=0$ everywhere outside the horizon so that $\phi$ carries no independent hair, whereas a neutron star's finite pressure and density source $\phi$ throughout the stellar interior, fixing the amplitude of the exterior Yukawa term.

The question turns out to hinge on a point that is easy to state once identified but is not obvious from the field equations alone: $f(R)=R+\alpha R^2$ gravity is dynamically equivalent, via the Legendre map $f'(R)-1$, to general relativity minimally coupled to a scalar field of mass $m_\phi^2=1/(6\alpha)$~\citep{Sotiriou2010,Sbisa2019}. A massive field mediates a force of finite range $\lambda=1/m_\phi=\sqrt{6\alpha}$, and its contribution to the metric of a static source is correspondingly a Yukawa-screened term, not a term proportional to $\alpha$ at fixed radius. This has an immediate consequence for any calculation, our own included, that begins from the ansatz $A(r)=1-2GM/r+\alpha\,a_1(r)+\mathcal{O}(\alpha^2)$ for the exterior metric: no such $a_1(r)$ exists, because $\lambda(\alpha)$ enters non-analytically. We derive the correct exterior solution below, and use it, rather than the power-series ansatz, for every calculation that follows. A second, related subtlety easy to miss is that the correct exterior solution is intrinsically isotropic and cannot be obtained by inverting the $g_{tt}$ potential to obtain $g_{rr}$ while leaving the angular part of the metric unscaled; we make this explicit in Sec.~\ref{sec:model} and verify it directly against the field equations.

Equipped with the correct exterior metric, we compute the weak-field deflection angle by the \GB\ theorem~\citep{Gibbons2008}, following the optical-metric construction that has by now been applied to a wide variety of compact-object spacetimes and extended to finite-distance source/observer configurations~\citep{Ishihara2016}, including other $f(R)$ black holes~\citep{Mandal2023,AparicioResco2026,Mohan2025} and $f(R)$ weak lensing by an independent, non-\GB\ route~\citep{Horvath2013}. The leading deflection angle is exactly the general-relativistic value, and the first $\alpha$-dependent correction is itself exponentially suppressed at astrophysical impact parameters. Four results follow and we believe the latter three to be new; the first being a new derivation of a cancellation already known from the post-Newtonian limit of analytic $f(R)$ gravity~\citep{Lubini_2011}. As for the new findings: first, a \GB\ optical-geometry derivation in which the cancellation of the scalaron is exact and transparent, rather than inferred from a fitted parameter. Second, the intrinsically isotropic gauge structure of the exterior, which the calculation turns out to require and which the customary areal substitution violates. Third, the explicit Yukawa coefficient of the first correction to the deflection angle, screened beyond the scalaron range and checked against direct quadrature. Fourth, the pressure-weighted neutron-star scalar charge, which has no analogue in the point-source treatment and which separates a neutron star from a black hole categorically.

Weak lensing is therefore closed as a channel once the impact parameter $b$ greatly exceeds $\lambda$, the surviving signatures lying either at sub-scalaron values of $b$ or in the strong-field, photon-sphere regime. We show, however, that the neutron-star exterior does retain a genuine, EOS-dependent handle on the theory through the pressure weighting of the scalar charge, absent for an idealised point mass, and we argue that this, in contrast to the deflection angle of an isolated black hole, is where a black-hole-versus-neutron-star comparison has most to offer.

The paper is organised as follows. Section~\ref{sec:model} sets out the two branches of the static, spherically symmetric vacuum solution and derives the Yukawa-screened exterior metric, including the gauge structure it requires. Section~\ref{sec:deflection} applies the \GB\ theorem to obtain the deflection angle. Section~\ref{sec:ns} extends the result to the neutron-star exterior and discusses the pressure-dependent scalar charge. Section~\ref{sec:results} presents the resulting figures. Section~\ref{sec:discussion} compares our results with the closely related recent literature and states what we believe to be original. Section~\ref{sec:conclusion} concludes. We adopt the signature $(-,+,+,+)$ and units $c=1$ throughout, restoring $G$ explicitly where it clarifies the physics.

\section{Model: two branches of the static exterior}
\label{sec:model}

For $f(R)=R+\alpha R^2$ the trace of the field equations is, using $f'''(R)=0$ identically,
\begin{equation}
6\alpha\,\Box R - R = 8\pi G T ,
\label{eq:trace}
\end{equation}
with $T=g^{\mu\nu}T_{\mu\nu}$ the trace of the matter stress-energy tensor.
Eq.~(\ref{eq:trace}) is the wave equation of a field of mass $m_\phi^2=1/(6\alpha)$, made explicit, following the original Legendre-map construction of the equivalence between quadratic gravity and general relativity plus a scalar field~\citep{Whitt1984}, by writing $f'(R)-1=2\alpha R \equiv \phi$,
\begin{equation}
\Box\phi - \frac{\phi}{6\alpha} = \frac{8\pi G}{3}\,T , 
\qquad m_\phi^2=\frac{1}{6\alpha}, 
\qquad \lambda \equiv \frac{1}{m_\phi} = \sqrt{6\alpha} .
\label{eq:scalaron}
\end{equation}

We adopt the general static, spherically symmetric line element with \emph{three} independent metric functions,
\begin{align}
ds^2 &= -A(r)\,dt^2 + B(r)\,dr^2 + D(r)\,r^2\,d\Omega^2 , \notag\\[2pt]
d\Omega^2 &\equiv d\theta^2 + \sin^2\theta\,d\varphi^2 ,
\label{eq:lineelement}
\end{align}
with $A(r)$, $B(r)$ and $D(r)$ independent~\citep{MTW1973}. Writing the angular part separately, rather than fixing $D(r) \equiv 1$ from the outset as is standard for a single-potential ansatz, is necessary: the two branches below occupy different corners of this three-function space, and collapsing prematurely to $D \equiv 1$ silently imposes on the scalaron branch an ansatz that it cannot satisfy.

\subsection{The Schwarzschild branch}

For the vacuum choice $A(r)=B(r)^{-1}=1-2GM/r$, $D(r)=1$ in Eq.~(\ref{eq:lineelement}), the Ricci scalar vanishes identically, so $R=0$ everywhere. Then $f'(R)=1$ and $f(R)=0$ identically, and the full field equation collapses to $R_{\mu\nu}=0$, already satisfied. Schwarzschild therefore solves the vacuum $f(R)=R+\alpha R^2$ field equations exactly, for every $\alpha$, with no scalaron hair. This is the branch on which most exact constructions of $f(R)$ black holes are built~\citep{Multamaki2006, delaCruzDombriz2009}, and it is the branch consistent with a horizon on which $\phi$ and its derivative vanish. Because this branch is an exact, nonlinear vacuum solution, the familiar areal-radius form ($D=1$) is entirely adequate here; the complication addressed below is specific to the branch sourced by matter.

\subsection{The scalaron-hair branch}

A star or a collapsing distribution of ordinary matter, however, sources $T\neq0$ throughout its interior, and Eq.~(\ref{eq:scalaron}) then forces $\phi\not\equiv0$ there; continuity of $\phi$ and of its normal derivative across the surface, the correct $f(R)$ junction condition~\citep{Deruelle2008}, which is strictly stronger than matching the metric and extrinsic curvature alone, propagates a nonzero $\phi$ into the exterior. This is the branch excited by the collapse or the presence of ordinary matter, and it is the one of physical interest here.

Linearizing the field equations about flat space with a static point source of mass $M$ and solving the resulting Helmholtz-type equations for the metric potentials $A(r)=1+2\Phi(r)$, $g_{ij}=(1-2\Psi(r))\delta_{ij}$ (isotropic gauge) gives
\begin{align}
\Phi(r) &= -\frac{GM}{r}\left[1+\frac13\,e^{-r/\lambda}\right], \notag\\[2pt]
\Psi(r) &= -\frac{GM}{r}\left[1-\frac13\,e^{-r/\lambda}\right],
\label{eq:potentials}
\end{align}
so that
\begin{equation}
A(r) = 1 - \frac{2GM}{r} - \frac{2GM}{3r}\,e^{-r/\lambda} ,
\qquad \lambda=\sqrt{6\alpha} .
\label{eq:metric}
\end{equation}

The exponential suppression of the scalaron outside the source is the same screening that protects $f(R)$ models against Solar-System and equivalence-principle bounds in the chameleon approach~\citep{Capozziello_2008}. There the suppression is density-driven, the scalaron mass growing in dense environments so that a spherical body acquires a thin shell; here the range $\lambda=\sqrt{6\alpha}$ is fixed by $\alpha$ alone. The isotropic-gauge construction that produces Eq.~(\ref{eq:potentials}) fixes $g_{ij}=(1-2\Psi(r))\delta_{ij}$ for the \emph{full} spatial metric, meaning that in the notation of Eq.~(\ref{eq:lineelement}),
\begin{equation}
B(r) = D(r) = C(r) \equiv 1-2\Psi(r) = 1+\frac{2GM}{r}-\frac{2GM}{3r}\,e^{-r/\lambda} ,
\label{eq:Cmetric}
\end{equation}
i.e.,\ the \emph{same} conformal factor $C(r)$ scales both the radial and angular parts of the spatial metric. This is a genuine physical constraint, not a coordinate convention, and it is not equivalent to the more familiar areal-radius substitution $B(r)=A(r)^{-1}$, $D(r)=1$ that is standard for single-potential solutions such as the Schwarzschild branch above. Expanding $A(r)^{-1}$ from Eq.~(\ref{eq:metric}) to linear order gives $A(r)^{-1}\approx1+2GM/r+(2GM/3r)e^{-r/\lambda}$: the Yukawa term carries the \emph{opposite sign} to the one appearing in the correct $C(r)$ of Eq.~(\ref{eq:Cmetric}). We verified directly, by substituting each candidate metric into the trace equation~(\ref{eq:trace}) and expanding to $\mathcal{O}(GM)$, that the areal-radius substitution ($B=A^{-1}$, $D=1$) leaves a nonzero residual in $(\Box-\lambda^{-2})R$ at this order, while the isotropic form of Eq.~(\ref{eq:Cmetric}) satisfies the linearised trace equation exactly. Eq.~(\ref{eq:Cmetric}) is therefore not an approximation to be refined, but the form required by the field equations at this order. The consequence is concrete: the isotropic form fixes the sign of the Yukawa term that enters the optical metric, and with it the sign of the deflection correction, so we carry Eq.~(\ref{eq:Cmetric}) through every calculation that follows.

One practical consequence is worth noting explicitly: because we work throughout in the isotropic radial coordinate, rather than converting to an areal one at an intermediate step, no additional coordinate-matching uncertainty of the kind that such a conversion would introduce enters the calculation. The radial coordinate $r$ appearing in Eqs.~(\ref{eq:potentials})--(\ref{eq:Cmetric}) is the same coordinate used consistently in Secs.~\ref{sec:deflection} and~\ref{sec:ns}.

The associated effective \PPN\ parameter, $\gamma(r)\equiv\Psi/\Phi= [1-\tfrac13 e^{-r/\lambda}]/[1+\tfrac13 e^{-r/\lambda}]$, interpolates between the Brans--Dicke, $\omega=0$, value $\gamma=1/2$ for $r\ll\lambda$ and the general-relativistic value $\gamma\to1$ for $r\gg\lambda$, in agreement with the standard scalar-tensor result~\citep{Sotiriou2010,Chiba2003,CapozzielloStabileTroisi2006}; we regard the recovery of this independently known limit as a useful check on Eqs.~(\ref{eq:metric})--(\ref{eq:Cmetric}), which we display in Fig.~\ref{fig:metric}. It is this same $\omega=0$ equivalence that underlies the Solar-System bound on $\alpha$: since $\gamma$ is constrained observationally to be within a part in $10^5$ of unity at Solar-System distances~\citep{Will2014}, $\lambda$ (and hence $\alpha$) must be far shorter than the scale of any such measurement, which is precisely the $r\gg\lambda$, screened regime relevant to the astrophysical impact parameters considered below.

It is worth being explicit about why Eq.~(\ref{eq:metric}) cannot be written as $A(r)=1-2GM/r+\alpha\,a_1(r)$ for any fixed function $a_1(r)$. The correction depends on $\alpha$ only through $\lambda=\sqrt{6\alpha}$, and $e^{-r/\sqrt{6\alpha}}$ has an essential singularity at $\alpha=0$ for every fixed $r>0$: all its Taylor coefficients in $\alpha$ vanish identically at $\alpha=0$, even though the function itself is not identically zero for $\alpha\neq0$. The correct small-coupling expansion is therefore an expansion in the weak-field amplitude $GM/r$ at fixed $\lambda$, not a Taylor expansion in $\alpha$ at fixed $r$; the two only coincide in the formal, and unphysical, limit $r\ll\lambda$ for every $r$ of interest simultaneously.

\section{Gauss--Bonnet deflection angle}
\label{sec:deflection}

Following \citep{Gibbons2008}, the deflection angle is obtained from the optical metric
\begin{equation}
d\bar{s}^2 = \frac{C(r)}{A(r)}\,dr^2 + \frac{C(r)\,r^2}{A(r)}\,d\varphi^2 ,
\label{eq:opticalmetric}
\end{equation}
with $C(r)$ given by Eq.~(\ref{eq:Cmetric}), obtained by setting $\theta=\pi/2$ and $ds^2=0$ in Eq.~(\ref{eq:lineelement}) with $B=D=C$. The deflection angle follows by integrating the Gaussian curvature $K_{\rm opt}$ of Eq.~(\ref{eq:opticalmetric}) over the domain $D_R$ bounded by the light ray of impact parameter $b$,
\begin{equation}
\hat\alpha = -\iint_{D_R} K_{\rm opt}\,dS .
\label{eq:GBintegral}
\end{equation}
Expanding $K_{\rm opt}$ to $\mathcal{O}(G^2M^2)$ we find that its leading, $\mathcal{O}(GM)$, piece carries no dependence on $\alpha$ whatsoever: this follows because $\Phi(r)+\Psi(r)=-2GM/r$ exactly, the scalaron terms of Eq.~(\ref{eq:potentials}) cancelling in the combination that controls null geodesics. Carrying out the integral of Eq.~(\ref{eq:GBintegral}) on this piece alone reproduces the standard result,
\begin{equation}
\hat\alpha^{(1)} = \frac{4GM}{b} ,
\label{eq:defl1}
\end{equation}
unmodified by $\alpha$ at this order, and we take the exact recovery of the coefficient $4$ as a check on the normalisation of our calculation.

The Yukawa piece of $K_{\rm opt}$ contributes only at $\mathcal{O}(G^2M^2/b^2)$ in amplitude; its full structure, $G^2M^2\lambda^{-1/2}b^{-3/2}e^{-b/\lambda}$, is given in Eq.~(\ref{eq:yukdefl}), and the two coincide only for $b\sim\lambda$.

A saddle-point evaluation of the resulting integral about the point of closest approach, valid for $b\gg\lambda$, gives
\begin{equation}
\hat\alpha_{\rm Yukawa}(b) \simeq
\frac{4\sqrt{2\pi}}{3}\,\frac{G^2M^2}{\sqrt{\lambda}\,b^{3/2}}\,
e^{-b/\lambda} , \qquad b\gg\lambda ,
\label{eq:yukdefl}
\end{equation}
so that the full expansion reads
\begin{align}
\hat\alpha(b) &= \frac{4GM}{b} + \frac{15\pi}{4}\frac{G^2M^2}{b^2} \notag\\[2pt]
&\quad + \frac{4\sqrt{2\pi}}{3}\,\frac{G^2M^2}{\sqrt{\lambda}\,b^{3/2}}\,
e^{-b/\lambda} + \mathcal{O}\!\left(\frac{G^3M^3}{b^3}\right) ,
\label{eq:fullexp}
\end{align}
the first term being the general-relativistic value recovered by the \GB\ construction~\citep{Gibbons2008}, the second the post-post-Newtonian coefficient of Refs.~\citep{EpsteinShapiro1980,FischbachFreeman1980}, and the third the genuinely new term. The provenance of the second term is worth recording. The coefficient $15\pi/4$ is the standard post-post-Newtonian result of Refs.~\citep{EpsteinShapiro1980,FischbachFreeman1980}; obtaining it from first principles would require the $\mathcal{O}(G^2M^2)$ pieces of the metric, which lie beyond the linear order retained here. The Yukawa coefficient $4\sqrt{2\pi}/3$ is correspondingly the estimate obtained from the leading-order metric, and we have verified it against direct quadrature rather than against a second-order calculation.

We have checked the coefficient $4\sqrt{2\pi}/3$ of Eq.~(\ref{eq:yukdefl}) against direct numerical quadrature of Eq.~(\ref{eq:GBintegral}), independent of the saddle-point expansion: the two agree to within $19\%$ at $b=4\lambda$, improving monotonically to within $7\%$ by $b=12\lambda$, consistent with the expected accuracy of a saddle-point approximation as it approaches its regime of validity $b\gg\lambda$.

The fractional deviation from general relativity,
\begin{equation}
\delta\hat\alpha_{\rm frac}(b) \equiv \frac{\hat\alpha_{\rm
Yukawa}(b)}{\hat\alpha^{(1)}(b)} = \frac{\sqrt{2\pi}}{3}\,\frac{GM}{\lambda}
\sqrt{\frac{\lambda}{b}}\;e^{-b/\lambda} ,
\label{eq:fracdev}
\end{equation}
falls exponentially with $b/\lambda$, as shown in Fig.~\ref{fig:deflection}; for $b\gtrsim10\lambda$ the deviation is already below one part in $10^4$ across the range $GM/\lambda\lesssim\mathcal{O}(1)$ displayed there. More generally, for $b\ge\lambda$ the fractional deviation falls below any chosen tolerance $\epsilon$ once $b/\lambda\gtrsim\ln[(GM/\lambda)/\epsilon]$, so that the screening sets in a few $e$-folds beyond the scalaron range, the mass entering only logarithmically and the reach of the effect being governed by $\lambda$.

For $b\lesssim\lambda$, where the exponential in Eq.~(\ref{eq:fracdev}) is no longer small, the saddle-point approximation breaks down and the deflection angle must instead be obtained by direct numerical integration of Eq.~(\ref{eq:GBintegral}) at fixed $b/\lambda$; this is also the regime in which the effective \PPN\ parameter $\gamma(r)$ of Sec.~\ref{sec:model} departs most strongly from unity, and where a strong-field, photon-sphere treatment such as that of Ref.~\citep{AparicioResco2026} is the more natural tool.

\section{Neutron star exterior}
\label{sec:ns}

A neutron star of finite radius $R_\NS$ sources $\phi$ throughout its interior according to Eq.~(\ref{eq:scalaron}), with the interior profile fixed by the same equation of state that determines the density and pressure profiles. Matching to the exterior at $r=R_\NS$ requires continuity of $\phi$ and $\phi'$, the $f(R)$ junction condition~\citep{Deruelle2008}, which is generally stricter than the general-relativistic Israel condition and has been implemented numerically, for realistic equations of state, by Refs.~\citep{Cooney2010,Yazadjiev2014,AstashenokCapozzielloOdintsov2014, CapozzielloDeLaurentisFarinelliOdintsov2016}, building on the earlier spherically symmetric $f(R)$ solutions of Ref.~\citep{CapozzielloStabileTroisi2008}. The same isotropic gauge structure established in Sec.~\ref{sec:model}, $B(r)=D(r)=C(r)$, applies here: the neutron-star exterior is not a Schwarzschild-like metric with a modified $g_{rr}$, but an isotropic one with a modified, pressure-weighted amplitude.

The resulting exterior again takes the form of Eq.~(\ref{eq:metric}), but with the amplitude of the Yukawa term set not by $GM/3$ alone, as for the idealised point source, but by a pressure-weighted integral of the trace of the stellar stress tensor~\citep{Sbisa2019},
\begin{equation}
A(r) = 1 - \frac{2GM}{r} - \frac{2Q_\phi}{r}\,e^{-r/\lambda} ,
\qquad Q_\phi \le \frac{GM}{3} ,
\label{eq:NSmetric}
\end{equation}
the equality being saturated only in the pressureless, point-mass limit. The bound holds for $R_\NS\ll\lambda$; when the stellar radius approaches the scalaron range the finite-size weighting of the interior integral competes with the pressure reduction, and $Q_\phi$ is no longer bounded above by $GM/3$. Setting $Q_\phi=GM/3$ in Eq.~(\ref{eq:NSmetric}) reproduces Eq.~(\ref{eq:metric}) term for term, so Eq.~(\ref{eq:metric}) is properly understood as the $Q_\phi\to GM/3$ special case of Eq.~(\ref{eq:NSmetric}) rather than as an independent result. Physically, the pressure that supports a neutron star against collapse partially cancels the contribution of its density to the scalar charge, so that a neutron star and a black hole of identical mass $M$ carry, in general, different values of $Q_\phi$. We parametrise this with $q\equiv3Q_\phi/GM\in[0,1]$, $q=1$ being the pressureless point-mass benchmark and $0<q<1$ a neutron star, without committing to a specific equation of state. It is worth being explicit that $q=1$ is \emph{not} the black-hole limit. The point source of Eq.~(\ref{eq:potentials}), with $T=-M\delta^3(\mathbf{x})$, is a pressureless \emph{matter} distribution; a static, isolated black hole is vacuum, occupies the $R=0$ branch, and carries no scalaron hair, so that $Q_\phi=0$ and $q=0$ exactly. This is the content of the no-hair results for scalar-tensor and $f(R)$ gravity~\citep{Hawking1972,SotiriouFaraoni2012,Canate2018,SultanaKazanas2018}, and it places the black hole at the opposite end of the range from the point-mass benchmark, rather than at the same end. Two qualifications belong with this statement. First, $Q_\phi$ is a source strength rather than a conserved charge: the scalaron is Yukawa-screened, so no conserved flux exists at infinity and $Q_\phi$ is a near-zone quantity fixed by the normalisation of Eq.~(\ref{eq:potentials}). Second, the result is specific to $f(R)=R+\alpha R^2$ and should not be read as a statement about quadratic gravity at large: uniqueness is contested in pure $R^2$~\citep{Kehagias2015}, and quadratic gravity including a Weyl term admits non-Schwarzschild black holes~\citep{Lu2015}, through a massive spin-2 mode that the present theory does not contain.

For a given equation of state, $q$ follows from the interior integration of Ref.~\citep{Cooney2010}, which lies outside the exterior-only treatment adopted here. Whether $q$ is sensitive to the equation of state at all remains open, since scalar charges in scalar-tensor gravity obey quasi-universal relations~\citep{YagiStepniczka2021}: a universal value would sharpen the prediction, a spread would tie it to mass-radius measurements such as those from NICER timing of PSR~J0740+6620~\citep{Raaijmakers2021}. The mass-radius relation of this same $f(R)=R+\alpha R^2$ theory has been computed for realistic equations of state (such as APR4, MPA1, SLy, WW1) in both metric and torsion formulations, with mass and compactness increasing with $\alpha$ in the metric case~\citep{Feola:2019zqg} --- this supplies the interior input against which such a comparison would be made.

Eqs.~(\ref{eq:fullexp}) and~(\ref{eq:NSmetric}) together imply that a black hole and a neutron star of equal mass in this theory are, at leading order, indistinguishable by weak lensing, since the leading term is $\alpha$-independent for either. At the next order, however, the distinction is not one of degree but of kind: the black hole, having $q=0$, carries no Yukawa term whatsoever and bends light exactly as general relativity requires, to all orders in $\alpha$, whereas a neutron star of the same mass retains a nonzero, pressure-weighted correction. The horizon-versus-surface distinction is therefore qualitative in this theory, rather than a matter of relative amplitude. We are careful not to overstate its observational reach: by Eq.~(\ref{eq:fracdev}) the neutron-star correction is itself exponentially small at $b\gg\lambda$, so that the comparison establishes a clean statement of principle and, in doing so, identifies the two channels, the photon sphere and the stellar interior, in which that principle may be turned into a measurement.

\section{Results}
\label{sec:results}

\begin{figure*}[htbp]
\centering
\includegraphics[width=\textwidth]{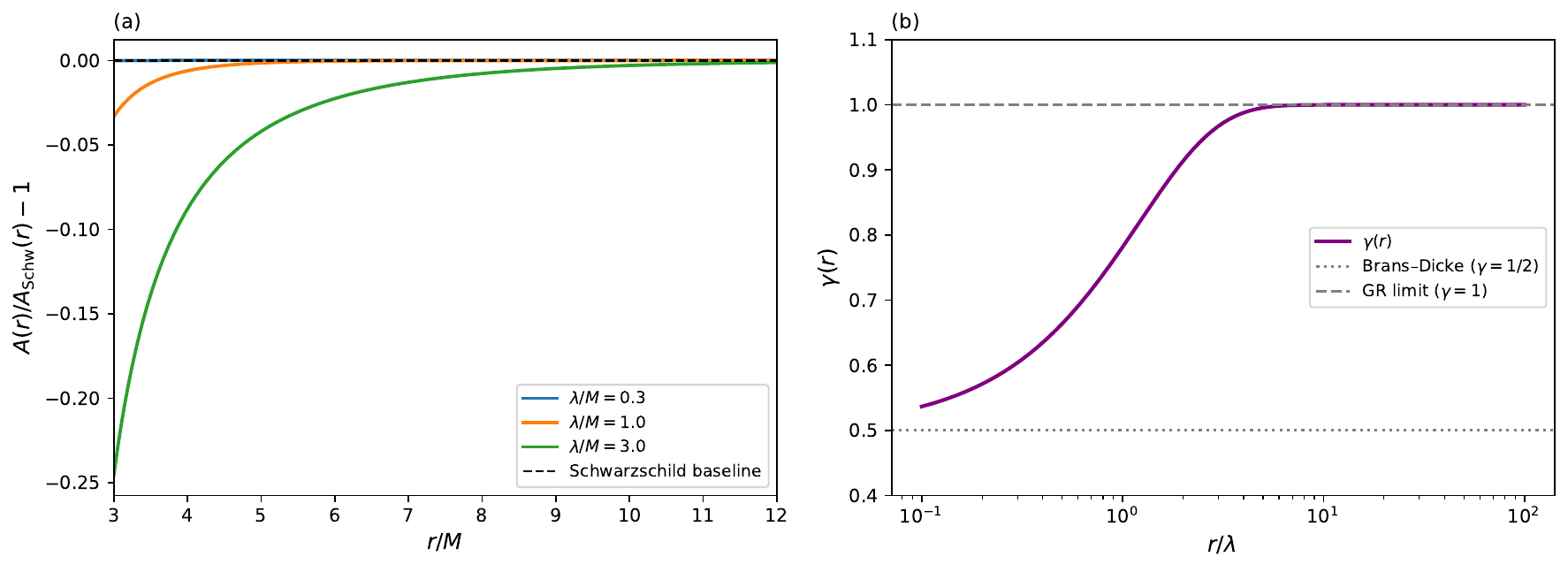}
\caption{(a) Exterior metric function $A(r)=-g_{tt}$ of Eq.~(\ref{eq:metric}) for $\lambda/M=0.3,1,3$, compared with Schwarzschild. (b) Effective \PPN\ parameter $\gamma(r)$, interpolating between the Brans--Dicke value $1/2$ at $r\ll\lambda$ and the general-relativistic value $1$ at $r\gg\lambda$, independent of $\lambda/M$ once plotted against $r/\lambda$. Panel~(a) is shown only over the range in which the linearised exterior solution is meaningful, the weak-field expansion having no validity at or inside the horizon, $r\le2M$.}
\label{fig:metric}
\end{figure*}

Figure~\ref{fig:metric} displays the exterior metric function $A(r)$ of Eq.~(\ref{eq:metric}) alongside the Schwarzschild curve, and the associated effective \PPN\ parameter $\gamma(r)$, illustrating the transition from the Brans--Dicke value $1/2$ at $r\ll\lambda$ to the general-relativistic value $1$ at $r\gg\lambda$.

\begin{figure*}[htbp]
\centering
\includegraphics[width=\textwidth]{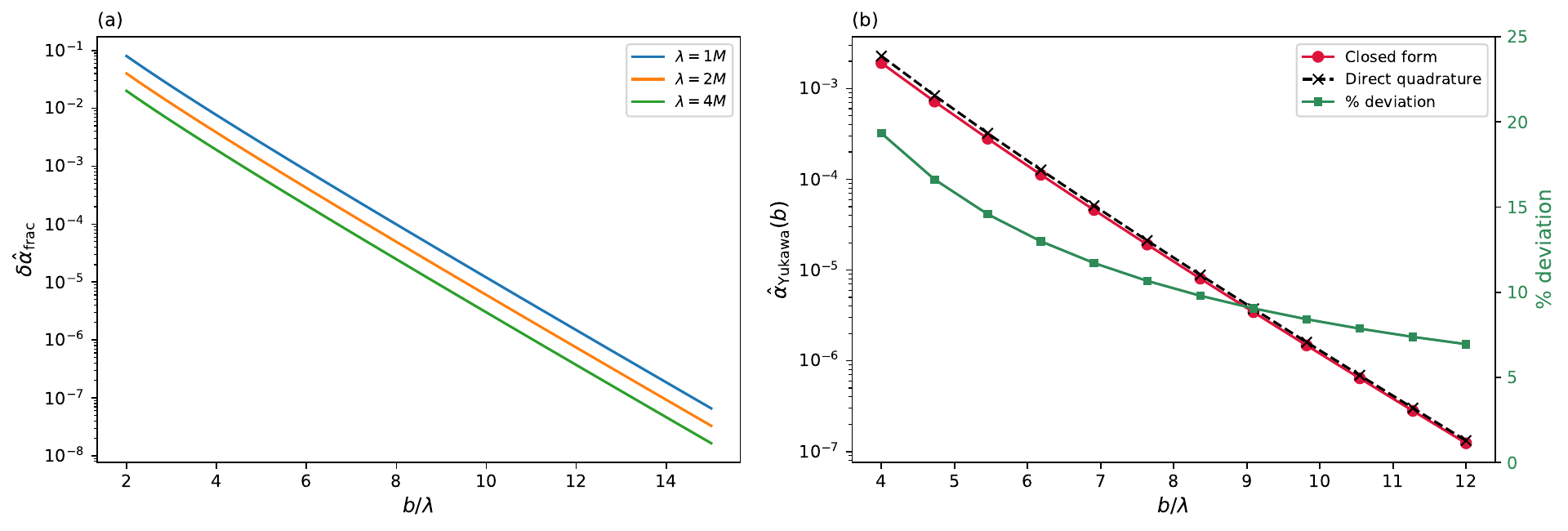}
\caption{Fractional deviation of the \GB\ weak-deflection angle from general relativity, Eq.~(\ref{eq:fracdev}), against impact parameter in units of the scalaron range $\lambda$, for three values of $\lambda/M$ (panel a). In panel~(b) the closed-form amplitude $\hat\alpha_{\rm Yukawa}(b)$ is cross-checked against direct numerical quadrature of Eq.~(\ref{eq:GBintegral}) (see Sec.~\ref{sec:deflection}). The saddle-point form systematically \emph{under}-estimates the quadrature, by $19\%$ at $b=4\lambda$ falling to $7\%$ at $b=12\lambda$ and thereafter as $\simeq(7/8)(\lambda/b)$, so the discrepancy is signed and monotonic rather than merely bounded, as expected of a saddle-point expansion approaching its regime of validity $b\gg\lambda$; this percentage difference is plotted on the right-hand axis of panel~(b).}
\label{fig:deflection}
\end{figure*}

Figure~\ref{fig:deflection} shows the fractional deviation of the deflection angle from general relativity, Eq.~(\ref{eq:fracdev}), as a function of $b/\lambda$ for several values of $\lambda/M$, making the exponential suppression explicit: the deviation falls by some six orders of magnitude across the interval $2\le b/\lambda\le15$ actually plotted, and by more than forty between $b=\lambda$ and $b=100\lambda$.

\begin{figure}[htbp]
\centering
\includegraphics[width=\columnwidth]{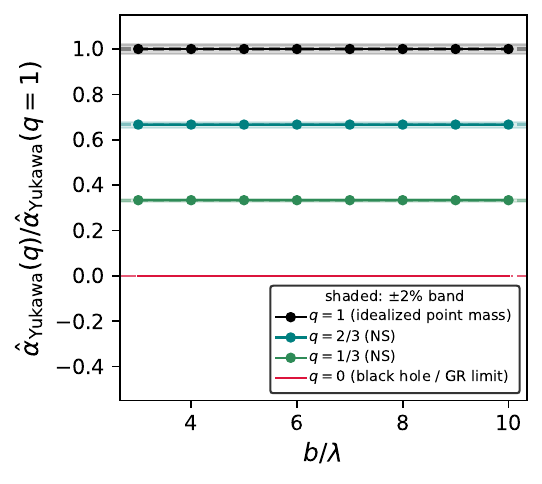}
\caption{Normalised Yukawa deflection angle ratio at fixed $\lambda = 1.5M$ across charge-to-mass parameters $q$. Numerical quadratures of Eq.~(\ref{eq:GBintegral}) for the pressureless point-mass benchmark ($q=1$) and neutron-star reduction factors ($q=2/3$ and $q=1/3$) are normalised to the $q=1$ curve. The ratios remain flat across $b/\lambda$ within the shaded $\pm 2\%$ bands around each reference value of $q$, confirming that the parameter $q$ acts purely as an overall amplitude rescaling of Eq.~(\ref{eq:NSmetric}) without altering the functional form of the deflection curve. The $q=0$ line corresponds to a static vacuum GR black hole, where the Yukawa contribution vanishes identically.}
\label{fig:bhns}
\end{figure}

Figure~\ref{fig:bhns} illustrates the Yukawa term of Eq.~(\ref{eq:NSmetric}) for a family of illustrative pressure-reduction factors $q$, a pure amplitude rescaling, as Eq.~(\ref{eq:NSmetric}) requires. We confirmed this independently, by computing each $q$ curve from its own numerical quadrature of Eq.~(\ref{eq:GBintegral}) rather than by rescaling a single reference curve.

\section{Discussion: relation to recent work}
\label{sec:discussion}

\begin{figure*}[htbp]
\centering
\includegraphics[width=\textwidth]{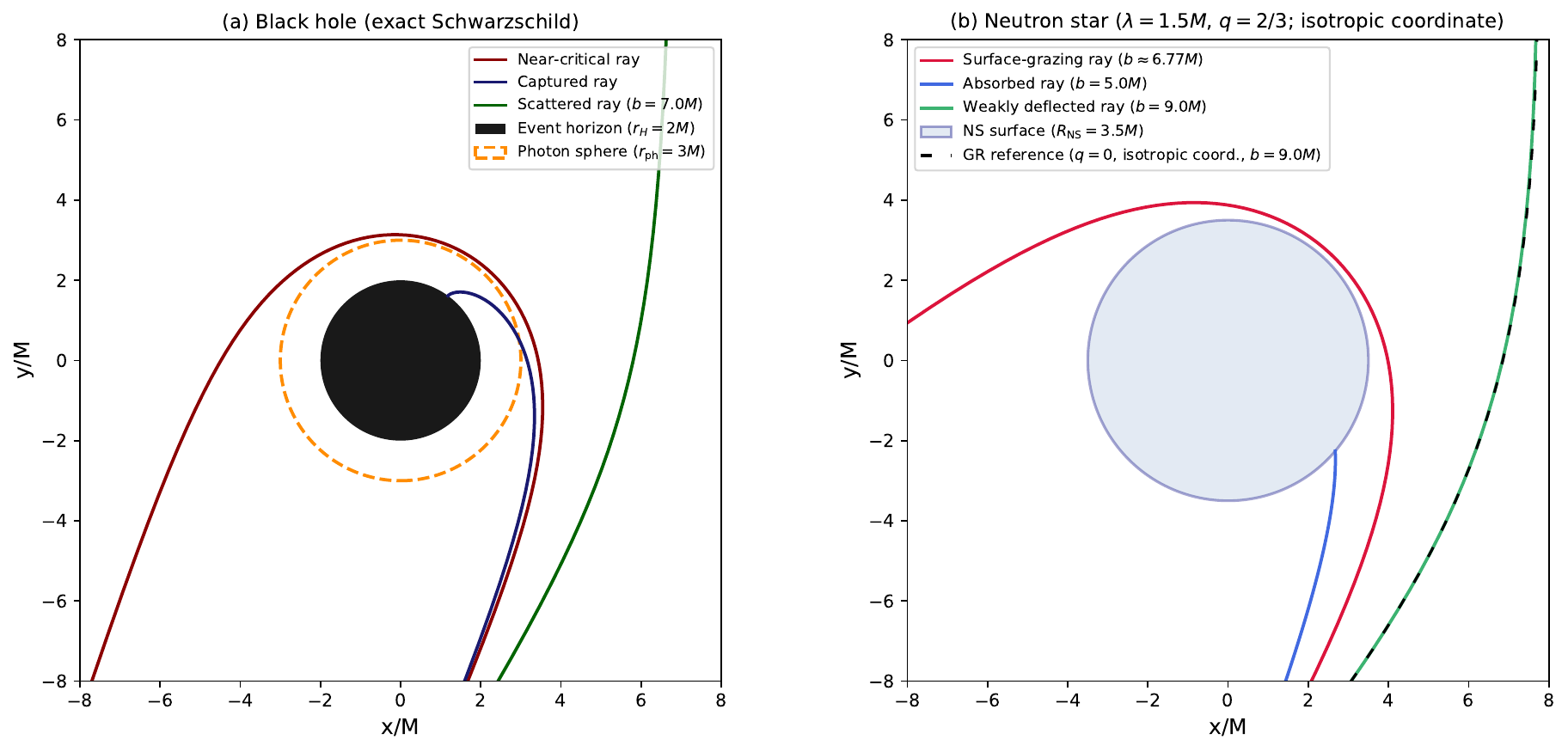}
\caption{Null geodesics in the two exteriors, at fixed $\lambda = 1.5 M$. (a) The black-hole branch, which is exact Schwarzschild for every $\alpha$, showing the horizon at $r_H=2M$, the photon sphere at $r_{ph}=3M$, and near-critical, captured, and scattered rays. (b) The neutron-star exterior of Eq.~(\ref{eq:NSmetric}) with $R_{NS}=3.5M$, for an illustrative reduction factor $q = 2/3$, drawn in the isotropic coordinate of Eq.~(\ref{eq:NSmetric}) and showing surface-grazing, absorbed, and weakly deflected rays; the dashed curve is the $q = 0$ (pure-GR) case of the same isotropic-coordinate family at the same impact parameter as the weakly deflected ray, which confirms that the two coincide once the Yukawa term is screened. The figure is illustrative rather than quantitative: the near-critical black-hole rays and the surface-grazing neutron-star ray both probe radii at which the linearised exterior is least secure, and $q$ here is not derived from an equation of state. Its purpose is to show where the two objects differ, namely at $b$ of order the horizon or stellar radius, and not at the astrophysical impact parameters of Fig.~\ref{fig:deflection}.}
\label{fig:rays}
\end{figure*}

Two recent papers address neighbouring ground on the black-hole side, and it is worth stating plainly how the present work differs from each. Mandal~\citep{Mandal2023} applies the \GW\ \GB\ theorem to a generic $f(R)$ black hole, computing the weak deflection angle together with the shadow and greybody factors, but does not adopt the specific $f(R)=R+\alpha R^2$ action nor identify the Yukawa structure of the correction. Aparicio Resco~\citep{AparicioResco2026} works with exactly this action and computes photon-sphere and strong-lensing observables from a numerically continued exterior solution, finding deviations of order several per cent to tens of per cent for parameters accessible to Event Horizon Telescope imaging; this is a strong-field result and is complementary to, not in tension with, our finding that the weak-field deflection deviation is exponentially small once $b\gg\lambda$. Read together, the two results delineate the theory's observational strategy: weak lensing is essentially blind to $\alpha$ at ordinary impact parameters, and the photon sphere, at $b$ comparable to $\lambda$ or to the horizon scale, is where the signal actually resides. Figure~\ref{fig:rays} makes this contrast
geometrically explicit, and displays in trajectory form the qualitative distinction argued for above: the black-hole exterior is exactly Schwarzschild, with no $\alpha$-dependence at
any order, whereas the neutron-star exterior carries the Yukawa term.

The cancellation itself was found first by Lubini \emph{et al.}~\citep{Lubini_2011}, who showed, for a general class of analytic $f(R)$ theories, that the Yukawa terms compensate in the post-Newtonian limit for massless particles, so that all lensing observables equal their general-relativistic values. Our result reproduces this within the quadratic model, and we make no claim of priority for it; what we add is its exhibition as an exact identity of the isotropic exterior and the closed-form screened term that survives it. A related comparison is owed to \citep{BerryGair2011}, who reached the equivalent conclusion in post-Newtonian language, both theories yielding an effective \PPN\ $\gamma=1$ over the regime probed. The two routes differ in what they establish. Theirs fixes a parameter and infers the absence of a signal; ours exhibits the cancellation directly, as the exact identity $\Phi(r)+\Psi(r)=-2GM/r$ in the combination that governs null geodesics, and supplies in closed form the screened term that survives it, Eq.~(\ref{eq:yukdefl}).

We note that while our exterior analysis assumes a vacuum background, real astrophysical environments with accretion structures or dark matter distributions restore the sourcing term $T_{\mathrm{matter}}$, exciting non-trivial scalar field profiles; we address this environmentally sourced scalaron, which is environmental and therefore leaves the no-hair theorem of Sec.~\ref{sec:ns} untouched \citep{Jacobson1999}, in a forthcoming paper.

\subsection{Observational channels and distinguishability}
\label{subsec:obs}

Observational probes naturally split along two distinct scale regimes. Relativistic images, photon rings, and logarithmic magnification near the shadow boundary belong exclusively to the strong-field, photon-sphere regime addressed by Ref.~\citep{AparicioResco2026}. Conversely, weak lensing at large impact parameters remains exponentially suppressed. For precision weak-field probes such as Shapiro time-delay measurements in tight pulsar binaries, the relevant benchmark is provided by high-precision timing observations, such as the Double Pulsar tests of general relativity by Kramer~\emph{et al.}~\citep{Kramer2021}. Pulsar observations have also been used to constrain $f(R)$ gravity directly, as in the analysis of SAX J1748.9-2021 by Nashed and Capozziello~\citep{nashed2024constrainingfcalrgravity}; such constraints are complementary to the propagation-based comparison made here. Comparing the fractional deflection deviation $\delta\hat\alpha_{\rm frac}$ of Eq.~(\ref{eq:fracdev}) against such observational thresholds underscores that constraints on $\alpha$ from weak-field propagation delays are far weaker than those available from strong-field imaging and from cosmology; a direct numerical comparison against the Double Pulsar's measured timing precision would quantify the margin, though the ordering of the channels is already clear without it. A third and complementary channel is X-ray pulse-profile modelling of the neutron-star surface itself: there the photons are emitted at $r=R_\NS$ and propagate outwards through the very exterior geometry derived here, so that the observed profile is sensitive to the metric between the surface and infinity independently of any background-source lensing. This route has been developed for extended matter distributions around neutron stars by Ref.~\citep{Shakeri:2022dwg}, and the same reasoning applies directly to the Yukawa term of Eq.~(\ref{eq:NSmetric}), whose amplitude it would probe at radii of order $R_\NS$, where the screening has not yet taken hold.

On the neutron-star side we are not aware of a prior application of the \GB\ deflection formalism to the $f(R)=R+\alpha R^2$ exterior matched to a physical stellar interior, nor of a prior statement that the resulting scalar charge, and hence the amplitude of the exponentially suppressed deflection correction, is pressure-weighted and therefore nonzero for a neutron star whilst vanishing identically for a black hole of equal mass. We regard this, together with the identification of the correct, Yukawa-screened, intrinsically isotropic exterior in place of a power-series ansatz, as the original content of the present work. The qualitative programme of asking whether horizon versus surface boundaries are separable through optical observables is of course shared with much of the compact-object gravity-testing literature~\citep{Vagnozzi2023,Biswas2024}; what is new here is the exterior solution on which the comparison rests, and the pressure-weighted charge that renders the comparison categorical rather than quantitative. Both the shadow observations underlying this literature, of M87$^\ast$~\citep{EHT2019M87} and of Sgr~A$^\ast$ itself~\citep{EHT2022SgrA}, and the broader goal of testing general relativity with compact-object observations more generally~\citep{Berti2015} and with pulsar timing in particular~\citep{Stairs2003}, provide the observational and conceptual backdrop against which the exponentially suppressed weak-lensing correction derived here should ultimately be judged.

\section{Conclusion}
\label{sec:conclusion}

We have shown that the static exterior of a matter source in quadratic $f(R)=R+\alpha R^2$ gravity is Schwarzschild plus a Yukawa term of range $\lambda=\sqrt{6\alpha}$, non-analytic in $\alpha$ at fixed radius, so that the customary power-series ansatz admits no solution at all. We further find the exterior to be intrinsically isotropic: a single conformal factor scales the radial and angular sectors alike, whereas the areal substitution standard for single-potential solutions reverses the sign of the Yukawa term and leaves a residual in the linearised trace equation. That substitution is habitual, and it fails silently.

Applying the \GW\ \GB\ theorem to this geometry, we recover the general-relativistic $4GM/b$ exactly, the scalaron cancelling in the combination that governs null geodesics. The first $\alpha$-dependent term is exponentially screened, and agrees with direct quadrature of the curvature integral to around $20\%$ at $b=4\lambda$, improving to $7\%$ at $b=12\lambda$. The optical-geometry route supplies the mechanism behind this cancellation, together with the screened coefficient in closed form; the resulting statement, that weak-field deflection alone cannot separate analytic $f(R)$ gravity from general relativity, is consistent with the post-Newtonian analysis of Ref.~\citep{BerryGair2011}.

Extending the calculation to a neutron-star exterior, we find that this is where the two objects part company, and that the separation is categorical. A static, isolated black hole is vacuum, carries no scalaron hair, and bends light exactly as general relativity requires, to all orders in $\alpha$. A neutron star of identical mass carries a scalar charge fixed by a pressure-weighted integral of its stress tensor, and does not. The horizon-versus-surface degeneracy that holds exactly in general relativity is therefore broken here, though by the presence or absence of a term rather than by its magnitude, and what survives is exponentially small.

Weak lensing is thus closed as a channel, and we are now in a position to say exactly why. The leverage passes instead to strong-field imaging at impact parameters of order the horizon scale, where the effective \PPN\ parameter departs most sharply from unity and next-generation interferometry will resolve photon-ring structure~\citep{AparicioResco2026}, and to the stellar interior, where the scalar charge is fixed by the equation of state and is thereby tied to the mass-radius and tidal measurements now becoming available. The horizon and the stellar surface are, in this theory, genuinely different objects. What remains is to measure the difference.

\section*{Acknowledgments}
Y.K. acknowledges that this research was supported by the Postdoctoral Fellowship Program (POSDOC) at the Universidad Nacional Aut\'onoma de M\'exico. I.L. thanks the Funda\c{c}\~ao para a Ci\^encia e Tecnologia (FCT), Portugal, for the financial support to the Centre for Astrophysics and Gravitation (CENTRA/IST/ULisboa) through grant No.~UID/PRR /00099/2025 (\doi{10.54499/UID/PRR/00099/2025}) and grant No. UID/00099/2025 (\doi{10.54499/UID/00099/2025}).

\bibliographystyle{iopart-num}
\bibliography{references}

\end{document}